\documentclass[aps,prl,twocolumn,showpacs,preprintnumbers,superscriptaddress,amsmath,amssymb]{revtex4}
\usepackage{graphicx}
\usepackage{dcolumn}
\usepackage{bm}
\usepackage{ulem}
\usepackage{color}

\newcommand{\s}{{\sigma}}

\def\be{\begin{eqnarray}}
\def\ee{\end{eqnarray}}

\newcommand{\nn}{\nonumber\\}

\renewcommand{\>}{\rangle}

\begin{document}

\title{The electronic instabilities of the Iron-based superconductors: a variational Monte-Carlo study}
\author{Fan Yang}
\affiliation{
Department of Physics,University of California at Berkeley,
Berkeley, CA 94720, USA}
\affiliation{Department of Physics, Beijing Institute of Technology, Beijing 100081, P.R.China}
\author{Hui Zhai}
\affiliation{Institute for Advanced Study, Tsinghua University, Beijing, 100084, China}
\author{Fa Wang}
\affiliation{Department of Physics, Massachusetts Institute of Technology, Cambridge, MA 02139}
\author{Dung-Hai Lee}
\affiliation{
Department of Physics,University of California at Berkeley,
Berkeley, CA 94720, USA}
\affiliation{Materials Sciences Division,
Lawrence Berkeley National Laboratory, Berkeley, CA 94720, USA}

\date{\today}

\begin{abstract}
We report the first variational Monte Carlo (VMC) study of the
iron-based superconductors. We use realistic band structures, and
the ordering instabilities/variational ansatzs are suggested by
previous functional renormalization group calculations. We examine
antiferromagnetism, superconducting pairing, normal state fermi
surface distortion and orbital order in the
 antiferromagnetic state.
\end{abstract}

\pacs{74.20.Mn, 74.72.-h, 74.25.Gz}
 \maketitle

The variational approach has a glorious history in condensed
matter physics. Examples of successful wavefunctions include the
BCS wave function, Laughlin wave function, Jastrow wave function,
{etc.}. In the study of the cuprate superconductivity, Anderson's
projected BCS wave function has been shown to capture many
important aspects of the cuprates\cite{rvbreview}. Despite the
successes, all variation approach have the draw back of being
biased: i.e., it bases on the assumption that the few variational
parameters built into the variational ansatz enable it to capture
the essence of the true ground state wavefunction.

Understanding the pairing mechanism and possible electronic
instabilities in iron-based
 superconductors have been a focus of interests in the past few years. This is not only for their high
 $T_c$, but also for many similarities they share with the cuprates.
Presently most of the works attribute the electron pairing to the
antiferromagnetic fluctuations, and it is widely believed that the
superconducting (SC) gap function takes opposite signs on the
electron and hole Fermi surfaces. This is partly because important
consequences of the opposite pairing sign predicted for the
neutron\cite{maier,korshunov} and STM exeriments\cite{wang,hu}
have now received experimental supports\cite{neutron,hanaguri}.

On the theory side, except Ref.\cite{seo} which assumes the
iron-based compounds are doped Mott insulator, most of the studies
are based on weak coupling
approximations\cite{mazin,kuroki,Li,FRG,chubukov,ikeda,scalapino,kuroki2,
maier2,zhai,FRG2,others}. Experimentally there are reports
favoring weak\cite{wlyang} and intermediate\cite{strong}
electronic correlation in this class of compounds. Among
Ref.\cite{mazin,kuroki,Li,FRG,chubukov,ikeda,scalapino,kuroki2,
maier2,zhai,FRG2,others} the functional renormalization group
(FRG)\cite{FRG} approach has the virtue of being unbiased. It sums
all one-loop particle-particle, particle-hole diagrams as well as
the vertex corrections. The results predict that in addition to
the AFM and SC instabilities, the iron-based compounds have
propensity toward Fermi surface distortion, magnetically coupled
orbital order, and charge density wave orderings\cite{zhai}.
However given the fact that there is an evidence that the
pnictides are intermediate coupling materials\cite{strong}, we
must ask which, if any, of the
 weak coupling results in Ref.\cite{mazin,kuroki,Li,FRG,chubukov,ikeda,scalapino,kuroki2,
maier2,zhai,FRG2,others} are valid. One of the main purpose of
this letter is to address the above question by performing a
variational Monte-Carlo (VMC) calculation. In addition, while FRG
only allows us to access the ordering tendencies, VMC allow us to
quantitatively study the ordered state. We use the realistic
bandstructures, and the variational ansatzs are guided by the
predictions of Ref.\cite{zhai}. We use partially projected wave
functions where both the degree of projection and the order
parameters are variational degrees of freedom.  We shall focus on
normal state Fermi surface distortion, AFM coupled orbital order,
and superconducting pairing. (In FRG the tendency toward charge
density wave order is the weakest.)

The band structure we use is that of Kuroki {\it et
al.}\cite{kuroki}. We model the electronic correlations by the
Hubbard and Hunds types of local interactions. The Hamiltonian is
given as follows:
\begin{eqnarray}
&&H=H_{\rm band}+
U_{1}\sum_{i\mu}n_{i\mu\uparrow}n_{i\mu\downarrow}+
U_{2}\sum_{i,\mu<\nu}n_{i\mu}n_{i\nu}+J_{H}\Big[\nn&&\sum_{\mu<\nu}
\sum_{\sigma\sigma^{\prime}}c^{+}_{i\mu\sigma}c^{+}_{i\nu\sigma^{\prime}}
c_{i\mu\sigma^{\prime}}c_{i\nu\sigma}+c^{+}_{i\mu\uparrow}c^{+}_{i\mu\downarrow}
c_{i\nu\downarrow}c_{i\nu\uparrow}+h.c.\Big]\label{H-H-model}
\end{eqnarray}
where $H_{\rm band}$ and its parameters can be found in
Ref.~\cite{kuroki} and its Table~I. In most of the paper we shall
use $\left(U_1,U_2,J_H\right)=\left(4,2,0.7\right)eV$. While these
parameters are compatible with Ref.\cite{strong} it is larger than
what's reported in Ref.\cite{wlyang}.The additional reasons for
doing so are (1) similar large correlation parameters are
concluded from LDA+DMFT calculations\cite{large} although with
controversery\cite{controversery}, (2) smaller interaction
parameters will make the energy gain associated with the
superconducting and orbital order too small for our VMC accuracy.
As the energy reduction due to normal state FS distortion is
relatively larger, smaller interaction parameters, e.g.
$\left(U_1,U_2,J_H\right)\sim\left(2,1,0.35\right)eV$, were also
used.

We use the following partially-projected variational
wave-functions,
\begin{equation}
\left|\psi\right>=g_{1}^{\hat{N}_{1}}g_{2}^{\hat{N}_{2}}\hat{P}_{N_{e}}\left|\psi_{MF}\right>,\label{trialwf}
\end{equation}
where
\begin{equation}
\hat{N}_{1}=\sum_{i,\mu}\hat{n}_{i\mu\uparrow}\hat{n}_{i\mu\downarrow},\quad\hat{N}_{2}=\sum_{i,\mu<\nu}\hat{n}_{i\mu}\hat{n}_{i\nu}.\label{gutzwiller}
\end{equation}
$\hat{P}_{N_{e}}$ fixes the total
number of electron to $N_e$, and $0\le g_{1},g_{2}\le 1$ suppress
configurations according to the number of electrons residing in
the same and different on-site orbitals. The mean-field wavefunction $|\psi_{MF}\>$
depends on the type of electronic order we are studying, and it will be discussed
in details later.

The VMC was carried out on $L^2=10\times10$ and $18\times18$
lattices with periodic boundary condition. In choosing the doping
level, we were careful  to avoid degeneracy in the band energy
between different Slater determinants. Standard Markovian chain
Monte Carlo approach was used with Metropolis update algorithm.
After an initial ``equillibration'' using $10^{5}$ Monte-Carlo
steps, about $10^{5},10^{6}$ and $10^{7}$ samples were used in the
study of Fermi surface distortion, orbital order and SC pairing
respectively. The adjacent samples are separated by $5L^{2}$
Monte-Carlo steps with each showing an acceptance ratio of about
0.23, which is enough to eliminate autocorrelation and thus
guaranty an efficient sampling. Due to the higher demand of
accuracy when studying SC pairing, we have applied the
``re-weighting'' scheme\cite{reweighting}. The error-bars are
estimated by calculating the variance of the energy expectation.

{\bf Normal state FS distortion} FRG predicts two leading FS
distortion: one preserves the $90^o$ rotation symmetry and the
other breaks it\cite{zhai}. The former shrinks both the electron
and hole pockets and produces a relative energy shift of the
electron and the hole bands\cite{donghui}. The later is the band
version of orbital ordering, and is suggested to be stabilized by
the AFM\cite{zhai}. The mean-field state $\left|\psi_{MF}\right>$
for the two types of FS distortion are the ground state of the
following quadratic Hamiltonian
\begin{eqnarray}
H_{MF}=\sum_{\mathbf{k}\alpha\sigma}(\varepsilon^{\alpha}_{\mathbf{k}}
+\chi({\bf k})) n_{\mathbf{k}\alpha\sigma}\label{MFPO}.
\end{eqnarray}
where $\varepsilon^{\alpha}_{\mathbf{k}}$ is the bare band
dispersion,
$n_{\mathbf{k}\alpha\sigma}=c^\dagger_{\mathbf{k}\alpha\sigma}c_{\mathbf{k}\alpha\sigma}$
and $\chi\left(\mathbf{k}\right)$ can be $\chi_{0}\cos k_x\cos
k_y$ ($90^o$ preserving) or $\chi_{0}\left(\cos k_x-\cos
k_y\right)$ ($90^o$ breaking).

\begin{figure}
\scalebox{0.8}{\includegraphics[scale=0.5]{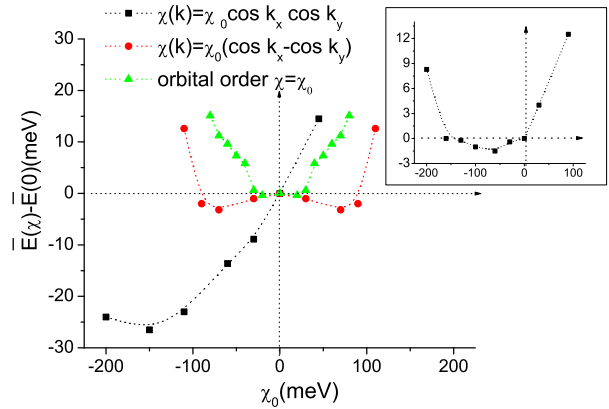}}
\caption{(Color online) Energy gain per site as functions of
$\chi_0$ for the normal state FS distortion, and the band and real
space version of orbital ordering. The results were obtained on a
$18\times18$ lattice under periodic boundary condition.  The
parameters $g_{1},g_{2}$ are fixed at their optimal values. The
size of points represents the error-bar and the dotted lines are
guide to the eyes. The main panel is for  $U_1=4, U_2=2,
J_H=0.7$eV at 0.6\% hole doping.  In the inset the result for at
13\% hole doping, with  $U_1=2, U_2=1, J_H=0.35$ eV and $\cos
k_{x}\cdot\cos k_{y}$ form-factor is shown.} \label{Fig.1}
\end{figure}

\begin{figure}
\scalebox{0.8}{\includegraphics[scale=0.8]{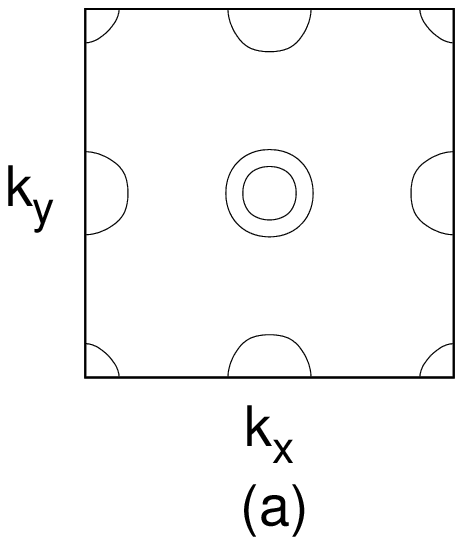}
\includegraphics[scale=0.8]{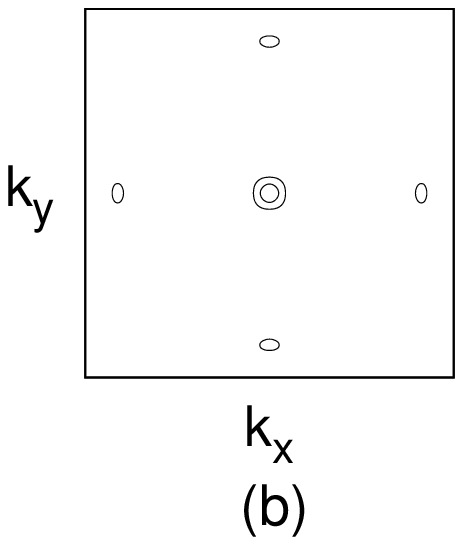}
} \caption{The Fermi surfaces at $0.6\%$ hole doping. (a) In the
absence of FS distortion. (b) The distorted Fermi surfaces
associated with $\chi \left(\mathbf{k}\right)=\chi_{0}\cos k_x\cos
k_y$.
 } \label{Fig.2}
\end{figure}

In the absence of AFM order, the results for $0.6\%$ hole-doping
on a $18\times18$ lattice using $(U_1,U_2,J_H)=(4,2,0.7)$eV  are
shown in the main panel of Fig.\ref{Fig.1}. The black and red
symbols represent $\bar{E}(\chi_0)-\bar{E}(0)$ as a function of
$\chi_0$ with $g_{1,2}$ fixed at their optimized values. (The
optimized $\left(g_{1},g_{2}\right)\approx
\left(0.31(4),0.76(4)\right)$ are nearly independent of $\chi$, as
the energy gain associated with their optimization is much larger
than that associated with optimization of $\chi$.) From
Fig.~\ref{Fig.1}, it is clear that while both $\cos k_x\cos k_y$
and $\cos k_x-\cos k_y$ FS distortions gain energy, the former is
the most energetically favorable one. This is consistent with the
FRG prediction. The green symbols in Fig.\ref{Fig.1} represent the
energy reduction due to a real space version of orbital order (see
later). We note that it is not energetically favorable in the
non-magnetic state. At the optimal $\chi_0$ the total energy gain
is about $27$meV per site. The negative $\chi_0$ moves the bands
near $\Gamma$ downward and the bands near $M$ upward. As a result
it shrinks both the electron and hole pockets (it actually splits
each electron pocket into two smaller ones). The distorted Fermi
surface is compared with the undistorted one in
Fig.~\ref{Fig.2}(a,b). The above results are completely consistent
with the FRG prediction\cite{zhai}. In the inset of
Fig.~\ref{Fig.1}, we show the $\bar{E}(\chi_0)-\bar{E}(0)$ versus
$\chi_0$ plot for $U_1=2, U_2=1, J_H=0.35$ eV with $\cos
k_{x}\cdot\cos k_{y}$ form-factor at 13\% hole doping. The
resulting relative energy shift of the bands near $\Gamma$ and $M$
is $90-126$ meV which is in good quantitative agreement with the
experimentally observed $\sim 100$ meV relative
shift\cite{donghui,quat_osci}.

\begin{figure}
\scalebox{0.75}{\includegraphics[scale=0.48]{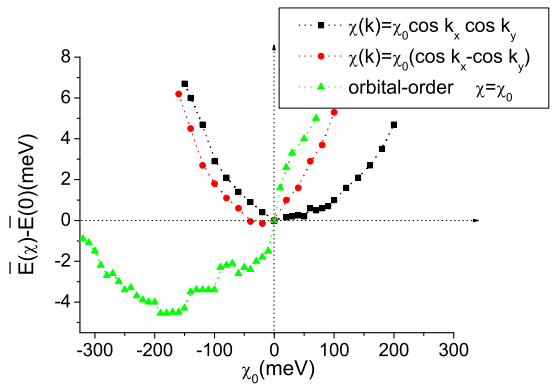}}
\caption{(Color online) The energy gain per site in the AFM state
as function of $\chi_0$ for the three types of electronic orders
studied in Fig.1(a). The same lattice size and interaction
parameters were used. The doping is 0.6\%-electron doping. The
variational parameters $g_{1},g_{2},M_{\mu\nu}$ are fixed at their
optimized values. The size of points represents the error-bar. The
dotted lines are just guide to the eyes.} \label{Fig.3}
\end{figure}

{\bf Orbital order in the AFM state} The $|\psi_{MF}\>$ for the AFM ordered state is the ground state of the
following mean-field
Hamiltonian,
\begin{eqnarray}
H_{MF}=H_0+\sum_{i\mu\nu\s}\left( \s
c^{+}_{i\mu\s}c_{i\nu\s}\right)M_{\mu\nu}e^{i\mathbf{Q}\cdot\mathbf{R}_{i}}\label{MFMAG},
\end{eqnarray}
where $H_0$ is given by either Eq. \ref{MFPO} or Eq. \ref{MFOR}.
Here $\mathbf{Q}=\left(\pi,0\right)$ is the AFM ordering wave
vector.  As for $M_{\mu\nu}$ we used the mean-field result of
Ref.\cite{Ying} where the non-zero $M_{\mu\nu}$ are $M_{\mu\mu},
\mu=1,..,5$,  and $M_{15}, M_{51}\left(=M_{15}\right)$. The
variational study is performed by keeping the ratio between the
optimal mean-field parameters
,$(M_{22},M_{33},M_{44},M_{55},M_{15})/M_{11}=(0.95, 1.03, 1.1,
0.76, -0.096)$, while allowing $M_{11}$ to vary. For
$(U_1,U_2,J_H)=(4,2,0.7)$eV at 0.6\% electron-doping (on $18\times
18$ lattice) we found
$\left(g_{1},g_{2}\right)=\left(0.49(0),0.76(6)\right),M_{11}=1.4eV$,
with a total energy reduction of about $3.0$eV per site, with
nearly $2\mu_B$ ordering moment. This value is significantly
larger than
 the measured ones for the stoichiometric compounds. The discrepancy can be due to
 the omission of the fluctuations in the orientation of magnetic moments and the ordering wavevectors ,
 and/or the large values of the interaction parameters.

In view of the strong atomic-like ordering moments in the AFM
state obtained above, when studying the orbital ordering in the
magnetic state we also adopt a real space version of  orbital
ordering\cite{Weiku}, where $|\psi_{MF}\>$ is the ground state of
\be
H_{MF}=\sum_{\mathbf{k}\alpha\sigma}\varepsilon^{\alpha}_{\mathbf{k}}
n_{\mathbf{k}\alpha\sigma}-\chi_0\sum_{i,\s}
(n_{i,yz\s}-n_{i,xz\s}) \label{MFOR} .\ee Here
$n_{i\alpha\sigma}=c^\dagger_{i\alpha\sigma}c_{i\alpha\sigma}$. We
note that the orbital $d_{xz},d_{yz}$ we use are $45^o$ rotated
from the $d_{XZ},d_{YZ}$ orbitals in Ref\cite{kuroki}. From Fig.3,
we conclude that in the presence of AFM, the real-space
$d_{xz}/d_{yz}$ orbital order is the most energetically favorable
(we suspect this is due to the fact that the large, localized,
ordering moment in the AFM state). It produces a total energy gain
of about $4.6$meV per site. At the optimal $\chi_0$ the
occupation-number difference between the $xz$ and $yz$ orbitals is
$ \bar{n}_{d_{xz}}-\bar{n}_{d_{yz}}=0.20(1). $ This value is
enhanced above the $\sim 0.15$ occupation difference already
present in the pure AFM  state. Our result agrees qualitatively
with a recent photoemission result\cite{lu} and a first principle
Wannier function calculation.\cite{Weiku}. Comparing this result
with the green symbols of  Fig.\ref{Fig.1}(a), we conclude that
this orbital order is stabilized by the AFM.

{\bf Superconducting pairing} The $\left|\psi_{MF}\right>$ we use
for studying pairing is the ground state of the following
mean-field Hamiltonian
\begin{eqnarray}
H_{MF}=\sum_{\mathbf{k}\alpha\sigma}(\varepsilon^{\alpha}_{\mathbf{k}}-\mu_c)
n_{\mathbf{k}\alpha\sigma}+\sum_{\bf
k\alpha}\left(\Delta^\alpha_{\bf k}
c^{+}_{\mathbf{k}\alpha\uparrow}c^+_{-\mathbf{k}\alpha\downarrow}+h.c.\right)\label{MFSC}.
\end{eqnarray}
We have studied four different types of gap functions:
\begin{eqnarray}
&&\Delta_{\mathbf{k}}^{\alpha}=\Delta_0\left(\cos k_{x}+\cos k_{y}\right),~\Delta_{\mathbf{k}}^{\alpha}=\Delta_0\cos k_{x}\cdot\cos k_{y}\nonumber\\
&&\Delta_{\mathbf{k}}^{\alpha}=\Delta_0,~~\Delta_{\mathbf{k}}^{\alpha}=\Delta_0\left(\cos k_{x}-\cos
k_{y}\right).\label{symmetry}
\end{eqnarray} In addition we only allow pairing in bands that cross the Fermi energy.
This is an excellent approximation because, as shown below, the gaps are quite small.

\begin{figure}
\scalebox{0.68}{\includegraphics[scale=0.38]{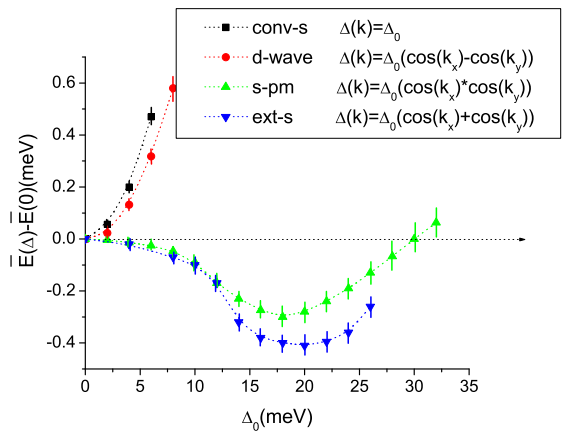}}
\caption{(Color online) Energy gain per site as function of
$\Delta_0$ for four different pairing form factors. The lattice
size is $10\times 10$, and $10\%$ hole doping  is studied. The
other variational parameters $g_{1},g_{2},\mu_{c}$ are fixed at
their optimized values. Error-bars are shown. The dotted lines are
guide to the eyes.}  \label{Fig.4}
\end{figure}

Our calculations for the SC state were carried out on a
$10\times10$ lattice at $10\%$ hole doping. The optimized
$\left(g_{1}, g_{2}\right)\approx\left(0.31(1), 0.76(2)\right)$.
Again, to an excellent approximation, the optimal
value of $g_{1,2}$ and $\mu_c$ do not depends on $\Delta_0$. The
results for $\bar{E}(\Delta_0)-\bar{E}(0)$ with $\left(g_{1}, g_{2},
\mu_{c}\right)$ fixed at their optimal values are shown in
Fig.~\ref{Fig.4}. The results suggest that while both
conventional-$s$ and $d_{x^2-y^2}$ pairing raise the energy,
$s_{\pm}$ and extended-$s$ lower it.
Interestingly, our result suggests the extended-$s$
($\Delta_0\left(\cos k_{x}+\cos k_{y}\right)$) form factor is
slightly  favored (by $0.1\pm 0.08$ meV) over the $s_{\pm}$
($\Delta_0\cos k_{x}\cdot\cos k_{y}$) one. On the surface this
contradicts the predictions of $s_\pm$ pairing form factor!
However it is important to realize that the $s_\pm$ gap function
is not synonymous to $\cos k_x\cos k_y$.  Indeed, the $s_\pm$ form
factor obtained from several weak coupling
approaches\cite{FRG,kuroki,scalapino} has strong variation around
the electron Fermi surfaces. We believe the near degeneracy of the
extended-$s$ and the $\cos k_x\cos k_y$ form factors suggests the
optimal pairing form factor is a linear combination of the two
(hence is anisotropic on the electron pockets). However it is
extremely computing time consuming to verify this, and we have not
been able to do it. In addition, we caution that the degree of the
gap function variation on the electron pockets will depends on the
values of the parameters. We cannot rule out that for other
parameter sets the $\cos k_x\cos k_y$ can be the leading pairing
form factor. Finally, the optimal gap amplitudes for the two
symmetries are about $18$meV and $20-22$meV respectively. Order of
magnitude wise these values are not far from the experimentally
measured gap scales.

In conclusion, we have performed variational Monte-Carlo
calculation to check the validity of our previous results based on
weak-coupling approximation. The results are qualitatively
consistent. We caution that because of the time consuming nature
of the calculation we are only able to study two sets of
interaction parameters. Clearly we can not rule of the possibility
that quantitative aspects of the above results will  be sensitive
to the precise values of the interaction parameters.

{\it Acknowledgement:} We thank Zheng-Yu Weng for sharing
computer resources and Hong Yao, Ying Ran and Tao Li for
helpful discussions. We acknowledge the support by the NSFC Grant
No.10704008 (FY); BRYS Program of Tsinghua University and NSFC Grant No. 10944002 (HZ); and the
DOE grant number DE-AC02-05CH11231 (DHL). This research
also use the resources of NERSC supported by the Office of Science of
the U.S. Department of Energy under Contract No.
DE-AC02-05CH11231.


\begin{thebibliography}{99}



\bibitem{rvbreview}
For a review and further references see P. W. Anderson, P. A. Lee, M. Randeria, T. M. Rice, N. Trivedi,
and F. C. Zhang, J. Phys.: Condens. Matter {\bf 16} R755 (2004).


\bibitem{maier} T.A.Maier, and D.J.Scalapino, Phys. Rev. B {\bf 78}, 020514(R) (2008).
\bibitem{korshunov} M.M. Korshunov, and I. Eremin, Phys. Rev. B {\bf 78}, 140509(R) (2008).
\bibitem{wang} F. Wang, H. Zhai, and D.-H. Lee, Europhys. Lett. {\bf 85}, 37005 (2009).

\bibitem{hu}Y.Y.Zhang, C.Fang, X. Zhou, K. Seo, W.F.Tsai, B.A. Bernevig, and J. Hu, Phys. Rev. B {\bf 80}, 094528 (2009).


\bibitem{neutron}
M. D. Lumsden {\it et al}, Phys. Rev. Lett. {\bf 102}, 107005
(2009); S.Chi {\it et al}, Phys. Rev. Lett {\bf 102}, 107006
(2009); S. Li {\it et al}, Phys. Rev. B {\bf 79}, 174527 (2009);
D. S. Inosov {\it et al}, Nature Physics 6, 178-181 (2010); J.
Zhao {\it et al}, arXiv:0908.0954 (2009).

\bibitem{hanaguri}
T. Hanaguri, S. Niitaka, K. Kuroki, and H. Takagi, Science {\bf
328}, 474 (2010).
\bibitem{seo} K.Seo, B. A. Bernevig, Jiangping Hu, Phys. Rev. Lett. {\bf 101}, 206404 (2008).


\bibitem{mazin} I.I. Mazin, D.J. Singh, M.D. Johannes, and M.H. Du, Phys. Rev. Lett. {\bf 101}, 057003 (2008).


\bibitem{Li} Z.J. Yao, J.X. Li, and Z. D. Wang, New J. Phys. {\bf 11}, 025009 (2009).


\bibitem{FRG}
 F. Wang, H. Zhai, Y. Ran, A. Vishwanath, and D.-H. Lee, arXiv:0805.3343 (2008);Phys. Rev. Lett. {\bf 102}, 047005 (2009).

\bibitem{chubukov}A. V. Chubukov, D. V. Efremov, and I. Eremin, Phys. Rev. B {\bf 78}, 134512 (2008).

\bibitem{ikeda}H. Ikeda, J. Phys. Soc. Jpn. {\bf 77}, 123707 (2008).

\bibitem{scalapino}
S. Graser, T. A. Maier, P. J. Hirschfeld, and D. J. Scalapino, New
J. Phys. {\bf 11}, 025016 (2009).

\bibitem{kuroki} K. Kuroki, S. Onari, R. Arita, H. Usui, Y. Tanaka, H. Kontani, and H. Aoki, Phys. Rev. Lett. {\bf 101}, 087004 (2008).

\bibitem{kuroki2} K. Kuroki, H. Usui, S. Onari, R. Arita, and H. Aoki, Phys. Rev. B {\bf 79}, 224511 (2009).

\bibitem{maier2} T.A. Maier, S. Graser, D.J. Scalapino, and P.J. Hirschfeld, Phys. Rev. B {\bf 79}, 224510 (2009).

\bibitem{zhai}H. Zhai, F. Wang, and D.-H. Lee, Physical Review B, {\bf 80}, 064517 (2009).

\bibitem{FRG2} F. Wang, H. Zhai, and D.-H. Lee, Phys. Rev. B {\bf 81}, 184512 (2010).

\bibitem{others}
A.V. Chubukov, M.G. Vavilov, and A.B. Vorontsov, Phys. Rev. B {\bf
80}, 140515(R) (2009); R. Thomale, C. Platt, J. Hu, C. Honerkamp,
and B. A. Bernevig, Phys. Rev. B {\bf 80}, 180505(R) (2009).

\bibitem{wlyang} W.L. Yang {\it et al}, Phys. Rev. B {\bf80}, 014508 (2009)
\bibitem{strong} See, e.g., M. M. Qazilbash, J. J. Hamlin, R. E.
Baumbach, Lijun Zhang, D. J. Singh, M. B. Maple, D. N. Basov
Nature Physics 5, 647 (2009).
\bibitem{large} L. Craco {\it et al}, Phys. Rev. B {\bf 78}, 134511 (2008); K. Haule {\it et al}, Phys. Rev. Lett. {\bf100}, 226402 (2008).

\bibitem{controversery} Some other LDA+DMFT studies listed below provide a
different picture for the model parameter. Namely, that U~4 is
relevant for a model including the As p-orbitals, but that a
reduced U should be used in the 5-band model. V. I. Anisimov, Dm.
M. Korotin, M. A. Korotin, A. V. Kozhevnikov, J. Kunes, A. O.
Shorikov, S. L. Skornyakov, S. V. Streltsov,  J. Phys.: Condens.
Matter 21 No 7, 075602(2009); M. Aichhorn, L. Pourovskii, V.
Vildosola, M. Ferrero, O. Parcollet, T. Miyake, A. Georges, and S.
Biermann, Phys. Rev. B 80, 085101(2009).


\bibitem{reweighting}
D. Ceperley, G. V. Chester, and K. H. Kalos, Phys. Rev. B {\bf
16}, 3081 (1977); C. J. Umrigar, K. G. Wilson, and J. W. Wilkins,
Phys. Rev. Lett. {\bf 60}, 1719 (1988).
\bibitem{donghui}
Donghui Lu, private communication.
\bibitem{lu}M. Yi, {\it et al},
arXiv 1011,0050.
\bibitem{Weiku} Chi-Cheng Lee, Wei-Guo Yin, and Wei Ku, Phys. Rev. Lett. 103, 267001 (2009)
\bibitem{quat_osci}
A.I. Coldea, J.D. Fletcher, A. Carrington, J.G. Analytis, A.F.
Bangura, J.-H. Chu, A.S. Erickson, I.R. Fisher, N.E. Hussey, R.D.
McDonald, Phys. Rev. Lett. 101, 216402 (2008).

 \bibitem{Ying} Ying Ran, Fa Wang, Hui Zhai, Ashvin Vishwanath, Dung-Hai
 Lee, Phys. Rev. B 79, 014505 (2009)


\bibitem{Fletcher} J.D. Fletcher, A. Serafin, L. Malone, J. Analytis, J-H Chu,
A.S. Erickson, I.R. Fisher, and A. Carrington, Phys. Rev. Lett.
{\bf 102}, 147001 (2009).

\bibitem{Hicks} C. W. Hicks, T. M. Lippman, M. E. Huber, J. G. Analytis, J. H. Chu,
A. S. Erickson, I. R. Fisher, and K. A. Moler, Phys. Rev. Lett.
{\bf 103}, 127003 (2009).



\end{thebibliography}
\end{document}